\documentstyle[11pt]{article}
\begin{document}
\begin{center}
{\bf QCD SCALES AND CHIRAL SYMMETRY IN FINITE NUCLEI}\\
\vspace{18pt}
D. G. Madland\\
\vspace{12pt}
Theoretical Division, Los Alamos National Laboratory\\Los Alamos, New
Mexico 87545, USA\\
\vspace{26pt}
{\bf Abstract}
\vspace{4pt}
\end{center}
\noindent
We report on our progress in the calculation of nuclear ground-state
properties using effective
Lagrangians whose construction is constrained by QCD scales and
chiral symmetry. Good evidence is found that QCD and chiral symmetry apply
to finite nuclei.\\
 
\noindent {\bf Introduction:} In 1992 a Dirac-Hartree calculation in mean field
approximation was performed by Nikolaus, Hoch, and Madland (NHM) [1] for the
nuclear ground-state properties of 57 nuclei and saturated nuclear matter.
Their
Lagrangian was motivated by empirically based improvements to the Walecka
scalar-vector model [2,3], but using contact interactions (point couplings) to
allow treatment of the Fock (exchange) terms. The nine coupling constants
of the
NHM Lagrangian were determined in a self-consistent procedure that solved the
model equations for several nuclei simultaneously in a nonlinear least-squares
adjustment algorithm with respect to well-measured nuclear ground-state
observables. The predictive power of the extracted coupling constants is better
than had been expected both for other finite nuclei and for the properties of
saturated nuclear matter.\\
 
\noindent In 1996 Friar, Madland, and Lynn (FML) [4] observed that whereas the
nine empirically based coupling constants of NHM span 13 orders of magnitude,
if they are instead scaled in accordance with the QCD-based Lagrangian of
Manohar and Georgi [5], and the role of chiral symmetry in weakening N-body
forces is taken into account (Weinberg [6-7]), then six of the nine scaled
coupling constants are {\it natural},
that is, they are dimensionless numbers of order 1.
This is potentially an important result because it may mean that (a) QCD and
chiral symmetry apply to finite nuclei and, if so, (b) heretofore unattainable
accuracy and predictive power in the nuclear many-body problem may be within
reach. Here, it is important to note that our work does not test QCD, or chiral
symmetry, but rather {\it effective Lagrangians} whose construction is
{\it constrained} by QCD and chiral symmetry.\\
 
\noindent In the next sections the NHM relativistic point coupling model is
briefly summarized, the roles of QCD scaling and chiral symmetry are briefly
discussed
and quantified, a more complete point coupling Lagrangian and first results
using it are presented, and the current status is given.\\
 
\noindent {\bf NHM Relativistic Point Coupling Model:}
The NHM model is a self-consistent Dirac Hartree-(Fock) model utilizing contact
interactions (point couplings) in the mean field ($\psi\;\longrightarrow\;
\langle
\,\psi\,\rangle$) and no Dirac sea approximations. The model consists of
four-,
six-, and eight-fermion point couplings leading to scalar and vector densities
with both isoscalar and isovector components, derivatives of the densities to
simulate the finite ranges of the mesonic interactions,\\
\newpage
\noindent but no explicit mean meson fields; instead, mean nucleon fields in
Skyrme-type approximation.
The Lagrangian is given by
 
\begin{equation}
{\cal L} = {\cal L}_{free} + {\cal L}_{4f} + {\cal L}_{hot} + {\cal L}_{der}
+ {\cal L}_{em} \ , \ \ {\rm where}
\end{equation}
 
\vspace{11pt}
\noindent ${\cal L}_{free}$ and ${\cal L}_{em}$ are the kinetic and
electromagnetic
terms, and
\begin{eqnarray}
{\cal L}_{4f} & = & -\frac{1}{2}{{\alpha}_{S}}({\bar{\psi}}{\psi})
({\bar{\psi}}{\psi})
-\frac{1}{2}{{\alpha}_{V}}({\bar{\psi}}{{\gamma}_{\mu}}{\psi})({\bar{\psi}}
{{\gamma}^{\mu}}{psi}) \nonumber \\
&   & -\frac{1}{2}{{\alpha}_{TS}}({\bar{\psi}}{\vec{\tau}}{\psi}){\cdot}
({\bar{\psi}}{\vec{\tau}}{\psi})
-\frac{1}{2}{{\alpha}_{TV}}({\bar{\psi}}{\vec{\tau}}{{\gamma}_{\mu}}{\psi})
{\cdot}
({\bar{\psi}}{\vec{\tau}}{{\gamma}^{\mu}}{\psi}) \ , \\
 & &  \nonumber \\
{\cal L}_{hot} & = & -\frac{1}{3}{{\beta}_{S}}({\bar{\psi}}{\psi})^{3}
-\frac{1}{4}{{\gamma}_{S}}({\bar{\psi}}{\psi})^{4} \nonumber \\
&   & -\frac{1}{4}{{\gamma}_{V}}[({\bar{\psi}}{{\gamma}_{\mu}}{\psi})({\bar
{\psi}}{{\gamma}^{\mu}}
{\psi})]^{2} \ , \ {\rm and} \\
 & &  \nonumber \\
{\cal L}_{der} & = & -\frac{1}{2}{{\delta}_{S}}({\partial_{\nu}}{\bar{\psi
}}\psi)
({\partial^{\nu}}{\bar{\psi}}\psi) \nonumber \\
&   & -\frac{1}{2}{{\delta}_{V}}({\partial_{\nu}}{\bar{\psi}}{\gamma_{\mu}}
\psi)
({\partial^{\nu}}{\bar{\psi}}{\gamma^{\mu}}\psi) \ .
\end{eqnarray}
 
\noindent In these equations, $\psi$ is the nucleon field, the subscripts
``$S$'' and ``$V$'' refer to the scalar and vector nucleon fields,
respectively, and the subscript ``$T$'' refers to isovector fields containing
the nucleon isospin ${\vec{\tau}}$. The physical makeup of ${\cal L}$ is that
${\cal L}_{4f}$ is a four-fermion interaction, while ${\cal L}_{hot}$ contains
six-fermion and eight-fermion interactions, and ${\cal L}_{der}$ contains
derivatives in the nucleon densities. There are a total of nine coupling
constants.\\
 
\noindent Minimizing the expectation value of the Hamiltonian corresponding to
Eq. (1) in the space of Slater determinants $|\phi\rangle$ leads to the
Dirac-Hartree
equations containing the following potentials:
\begin{eqnarray}
{V_{S}} & = & {\alpha_{S}}{\rho_{S}} + {\beta_{S}}{\rho_{S}^{2}} + {\gamma_{S
}}{\rho_
{S}^{3}} + {\delta_{S}}{\Delta}{\rho_{S}} \ , \\
{V_{V}} & = & {\alpha_{V}}{\rho_{V}} + {\gamma_{V}}{\rho_{V}^{3}} + {\delta_{V}}
{\Delta}{\rho_{V}} \ , \\
{V_{TS}} & = & {\alpha_{TS}}{\rho_{TS}} \ , \ {\rm and} \\
{V_{TV}} & = & {\alpha_{TV}}{\rho_{TV}} \ ,
\end{eqnarray}
 
\noindent where Eq. (5) is the isoscalar-scalar potential corresponding to
$\sigma$ meson (fictitious) exchange, Eq. (6) is the isoscalar-vector potential
corresponding to $\omega$ meson exchange, Eq. (7) is the isovector-scalar
potential corresponding to $\delta$ meson exchange, and Eq. (8) is the
isovector-vector potential corresponding to $\rho$ meson exchange.
In these latter equations the scalar density is given by $\rho_{S}\;=\;\langle
\phi|{\bar{\psi}}{\psi}|\phi\rangle$, the vector density is given by $\rho_{V}
\;=\;\langle\phi|{\bar{\psi}}{\gamma_{0}}{\psi}|\phi\rangle$, the
isovector-scalar
density is given by $\rho_{TS}\;=\;\langle\phi|{\bar{\psi}}
{\tau_{3}}{\psi}|\phi\rangle$,
and the isovector-vector density is given by $\rho_{TV}\;=\;\langle\phi|
{\bar{\psi}}{\tau_{3}}{\gamma_{0}}{\psi}|\phi\rangle$.\\
 
\noindent The nine coupling constants of the NHM Lagrangian were determined in
a self-consistent procedure that solves the Dirac-Hartree equations for
several nuclei simultaneously in a nonlinear least-squares adjustment
algorithm of Levenberg-Marquardt type with respect to well-measured nuclear
ground-state observables.
The well-measured observables used are (a) the \\
\newpage
\noindent ground-state masses (binding energies),
(b) the rms charge radii, and (c) the spin-orbit splittings of the least-bound
neutron and proton spin-orbit pairs.
The spherical closed-shell nuclei $^{16}$O, $^{88}$Sr, and $^{208}$Pb were
chosen for the determination of the coupling constants (12 observables to
determine 9 coupling constants).
The NHM coupling constants are given in Table 1 where the first four
coupling constants refer to Eq. (2), the next three refer to Eq. (3), and
the remaining two refer to Eq. (4).
They span 13 orders of magnitude.
 
\begin{table} [h]
\centering
\caption{Optimized Coupling Constants for the Relativistic Point Coupling
Model}
\vspace{11pt}
\begin{tabular}{|c|r|c|} \hline
Coupling Constant & Magnitude & Dimension \\ \hline
$\alpha_{S}$ & -4.508${\times}10^{-4}$ & MeV$^{-2}$ \\
$\alpha_{TS}$ & 7.403${\times}10^{-7}$ & MeV$^{-2}$ \\
$\alpha_{V}$ & 3.427${\times}10^{-4}$ & MeV$^{-2}$ \\
$\alpha_{TV}$ & 3.257${\times}10^{-5}$ & MeV$^{-2}$ \\
$\beta_{S}$ & 1.110${\times}10^{-11}$ & MeV$^{-5}$ \\
$\gamma_{S}$ & 5.735${\times}10^{-17}$ & MeV$^{-8}$ \\
$\gamma_{V}$ & -4.389${\times}10^{-17}$ & MeV$^{-8}$ \\
$\delta_{S}$ & -4.239${\times}10^{-10}$ & MeV$^{-4}$ \\
$\delta_{V}$ & -1.144${\times}10^{-10}$ & MeV$^{-4}$ \\ \hline
\end{tabular}
\end{table}
 
\noindent With these nine coupling constants one can calculate the following
for spherical
closed-shell nuclei: (a) single-particle Dirac wave functions and eigenvalues
for both protons and neutrons, (b) nuclear ground-state mass and binding
energy,
(c) proton and neutron densities and their moments, (d) nuclear charge density
and its moments, and (e) isoscalar- and isovector-, scalar and vector,
potentials.
For example, Table 2 gives the average absolute deviations of calculated
binding
energies and rms charge radii from the measured values for a number of cases.
This is an encouraging result, especially if one notes that the corresponding
rms deviations are even smaller.
 
\begin{table} [h]
\centering
\caption{Average Absolute Deviations of Calculated Observables from Measured
Observables for the Relativistic Point Coupling Model}
\vspace{11pt}
\begin{tabular}{|c|c|c|} \hline
Observable & Avg. Abs. Deviation & Number of Cases \\ \hline
$E_{B}$ & 2.52 MeV & 34 \\
$<r^{2}>_{charge}^{1/2}$ & 0.020 fm & 17 \\ \hline
\end{tabular}
\end{table}
 
\noindent {\bf QCD Scales and Chiral Symmetry:}
An $SU(2) \times SU(2)$ Lie algebra is generated by the commutation rules
of vector and
axial charges. Assuming that axial currents are approximately conserved,
the resulting
symmetry is called {\it chiral symmetry.} In the exact chiral limit quarks
are massless
and the Goldberger-Treiman relation, connecting the strong and weak
interactions, is exact:
\begin{equation}
g_{A}(0) = \frac{G_{A}}{G_{V}} = g_{\pi{NN}} \frac{f_{\pi}}{m_{N}} \; ,
\end{equation}
where $g_{A}$ is the strong axial-vector coupling constant, $G_{A}$ and
$G_{V}$ are
the weak axial-vector and polar-vector coupling constants, respectively, $g
_{\pi{NN}}$ is
the effective pion-nucleon coupling constant, $f_{\pi}$ is the pion decay
constant, and
$m_{N}$ is the nucleon mass. In 1990 Weinberg [6] \\
\newpage
\noindent addressed the (broken) chiral symmetry and introduced chiral
perturbation theory into nuclear physics and
showed that chiral Lagrangians predict the suppression of N-body forces.
He accomplished this by constructing the most general possible chiral
Lagrangian
involving pions and low-energy nucleons as an infinite series of allowed
derivative
and contact interaction terms and by using QCD mass scales and dimensional
power counting to categorize
the terms of the series according to their characteristic (average) momentum
or
energy scales.
He concluded that N-body forces were a series in the ratio of a small
momentum scale
to a large one, leading to a systematic suppression of N-body forces.
That is, the infinite series is not physically infinite.\\
 
\noindent Consider the generic Lagrangian for pions ($\vec{\pi}$) and
nucleons ($\psi$) and containing derivatives, ($\partial^{\mu}$), used in
dimensional power counting by Manohar and Georgi [5], and later refined by
Weinberg [7] and Lynn [8]:
\begin{equation}
{\cal L} \sim -c_{l m n}
\left[ \frac{\overline{\psi}\psi}{f^2_{\pi} \Lambda} \right]^l
\left[ \frac{\vec{\pi}}{f_{\pi}} \right]^{m}
\left[ \frac{\partial^{\mu}, m_{\pi}}{\Lambda} \right]^n
f^2_{\pi} \, \Lambda^2 \,
\end{equation}
where $f_{\pi}$ and $m_{\pi}$ are the pion decay constant, 92.4 MeV, and pion
mass, 139.6 MeV, respectively. If the theory is {\it natural} [5,8], this
Lagrangian should lead to dimensionless
coefficients $c_{lmn}$ of order unity for each order in the QCD large-mass
scale, $\Lambda$ = 1 GeV. The chiral constraint is given by
\begin{equation}
\Delta = l + n - 2 \geq 0,
\end{equation}
which guarantees that no $\Lambda$ appears in the numerator of $\cal L$,
and hence
the Lagrangian is a series in $\Lambda^{-1}$ and therefore converges.
Thus, all information on scales ultimately resides in the $c_{lmn}$. If they
are natural, QCD scaling works. \\
 
\noindent As a first test, the nine coupling constants of the NHM Lagrangian
are again shown in Table 3,
both in dimensional (identical to Table 1) and dimensionless form, the latter
obtained by equating Eqs. (2)--(4) and Eq. (10).
[Note that we use ${\vec{\tau}}$ in Eq. (2) and ${\vec{t}}\;=\;\frac{1}{2}
{\vec{\tau}}$ in Eq. (10)]. One sees that the nine terms of the NHM
Lagrangian represent portions of three different orders in the large-mass
QCD scale $\Lambda$, namely, $\Lambda^{0}$, $\Lambda^{-1}$, and $\Lambda^{-2}
$.
However, while the nine
\begin{table} [h]
\centering
\caption{Optimized Coupling Constants for the Relativistic Point Coupling
Model and Corresponding Dimensional Power Counting Coefficients
and Chiral Expansion Order}
\vspace{11pt}
\begin{tabular}{|c|r|c|c|l|} \hline
Coup. Const. & Magnitude & Dimension & \mbox{\boldmath$c_{lmn}$} & Order \\
\hline
$\alpha_{S}$ & -4.508${\times}10^{-4}$ & MeV$^{-2}$ & -1.98 & $\Lambda^{0}$
\\
$\alpha_{TS}$ & 7.403${\times}10^{-7}$ & MeV$^{-2}$ & 0.0128 & $\Lambda^{0}$
\\
$\alpha_{V}$ & 3.427${\times}10^{-4}$ & MeV$^{-2}$ & 1.48 & $\Lambda^{0}$ \\
$\alpha_{TV}$ & 3.257${\times}10^{-5}$ & MeV$^{-2}$ & 0.56 & $\Lambda^{0}$
\\
$\beta_{S}$ & 1.110${\times}10^{-11}$ & MeV$^{-5}$ & 0.28 & $\Lambda^{-1}$
\\
$\gamma_{S}$ & 5.735${\times}10^{-17}$ & MeV$^{-8}$ & 9.28 & $\Lambda^{-2}$
\\
$\gamma_{V}$ & -4.389${\times}10^{-17}$ & MeV$^{-8}$ & -7.10 & $\Lambda^{-2}$
\\
$\delta_{S}$ & -4.239${\times}10^{-10}$ & MeV$^{-4}$ & -1.84 & $\Lambda^{-2}$
\\
$\delta_{V}$ & -1.144${\times}10^{-10}$ & MeV$^{-4}$ & -0.49 & $\Lambda^{-2}$
\\ \hline
\end{tabular}
\end{table}
\newpage
\noindent original coupling constants span 13 orders of magnitude,
the dimensional-power-counting coefficients $c_{lmn}$ are six of order (1),
two of order (10), and one of order (10$^{-2}$).
Given that the $c_{lmn}$ should be of order unity if they are natural this is
a surprisingly good result since it has been obtained with an incomplete mix
of terms from three orders in $\Lambda$ and the pions have been ignored,
that is, assumed to cancel out. Presumably, the absent terms are represented
by unphysical (unnatural) values of some of the existing $c_{lmn}$ and this
introduces yet other unphysical consequences. The question is can we improve
on this situation? \\
 
\noindent {\bf Revisit the Relativistic Point Coupling Model:}
In our subsequent calculations studying the NHM coupling constants we
observed that while only one
of the two isospin-dependent terms is natural, namely, $c_{\alpha_{TV}}$,
the sum of the two, namely, ($c_{\alpha_{TS}}\:
+\:c_{\alpha_{TV}}$), appears to be natural. [We connect the 1st
and 4th columns of Table 3 with this notation].
We also observed that while both eight-fermion interactions are unnatural,
their sum ($c_{\gamma_{S}}\:+\:c_{\gamma_{V}}$) appears to be natural.
These two behaviors persist throughout an exploration of
the $\chi^{2}$ space occupied by the nine coupling constants of the NHM
Lagrangian. Therefore, we performed a more detailed grid-search on
$c_{\alpha_{TS}}$ and $c_{\alpha_{TV}}$ that resulted in a total
of three separate minima corresponding to three different sets of these
coupling constants. These are, approximately, the sets
\{0.01,0.56\} (original NHM set), \{-0.37,0.92\}, and \{-1.50,1.88\}.
In the latter two cases the individual coupling constants
are natural and their sums are natural whereas in the former case only
one coupling constant is natural, but the sum is natural.
Note that the three sums are similar having the values, respectively,
of 0.57, 0.55, and 0.38. The calculations corresponding to these three
minima are labeled (a), (b), and (c), respectively, in Table 4.
[The * appearing for calculation (a), that of the original NHM set, refers
to the fact that the spin-orbit pair weights here are more stringent than
for the
remainder of the table so the $\chi^{2}$/pt is correspondingly higher].
In this table, $\langle\delta\rm{BE}\rangle$ and $\langle\delta\rm{RMS}
\rangle$
are the average absolute deviations of calculated from measured binding
energies
and rms charge radii, respectively (as in Table 2).
 
\begin{table} [h]
\centering
\caption{Summary of Important Calculations to Date}
\vspace{11pt}
\begin{tabular}{|c|c|c|r|c|c|} \hline
Calc. & \# Coup. Const. & \# Natural & $\chi^{2}$/pt & $\langle\delta\rm{BE
}\rangle$
& $\langle\delta\rm{RMS}\rangle$ \\
 & & & & (MeV) & (fm) \\ \hline
a & 9 & 6 & $^{*}$8.94 & 2.52 & 0.020 \\
b & 9 & 7 & 4.10 & 2.86 & 0.021 \\
c & 9 & 7 & 4.33 & 3.78 & 0.031 \\ \hline
d & 10 & 8 & 4.94 & 2.60 & 0.022 \\ \hline
e & 9 & 9 & 57.61 & 8.90 & 0.125 \\
f & 10 & 10 & 68.92 & 6.31 & 0.118 \\ \hline
\end{tabular}
\end{table}
 
\noindent Calculations (b) and (c), in comparison to (a), show that although
naturalness occurs for 7 of 9 coupling constants in two cases, neither of
these has better predictive power than the original NHM set with only 6
natural coupling constants.
However, it is clear that the predictive power of calculation (b) is only
slightly less good than that of calculation (a). \\
 
\noindent These results suggest that the incompleteness of the NHM
Lagrangian (four terms of order $\Lambda^{0}$, one term of order
$\Lambda^{-1}$, and four terms of order $\Lambda^{-2}$) should be
addressed. In particular, there are only two terms containing isospin
(both of order $\Lambda^{0}$) and only one term of order $\Lambda^{-1}$. \\
\newpage
\noindent Accordingly, a new term was added to the Lagrangian, with coupling
constant $\beta_{TS}$, that is of isovector-scalar character and of order
$\Lambda^{-1}$. The new Lagrangian is given by
 
\begin{eqnarray}
{\cal L}_{4f} & = & -\frac{1}{2}{{\alpha}_{S}}({\bar{\psi}}{\psi})({\bar{
\psi}}{\psi})
-\frac{1}{2}{{\alpha}_{V}}({\bar{\psi}}{{\gamma}_{\mu}}{\psi})({\bar{\psi}}
{{\gamma}^{\mu}}{\psi}) \nonumber \\
&   & -\frac{1}{2}{{\alpha}_{TS}}({\bar{\psi}}{\vec{\tau}}{\psi}){\cdot}({\
bar{\psi}}{\vec{\tau}}{\psi})
-\frac{1}{2}{{\alpha}_{TV}}({\bar{\psi}}{\vec{\tau}}{{\gamma}_{\mu}}{\psi})
{\cdot}
({\bar{\psi}}{\vec{\tau}}{{\gamma}^{\mu}}{\psi}) \ , \\
 & &  \nonumber \\
{\cal L}_{hot} & = & -\frac{1}{3}{{\beta}_{S}}({\bar{\psi}}{\psi})^{3}
-\frac{1}{3}{{\beta}_{TS}}({\bar{\psi}}{\vec{\tau}}{\psi}){\cdot}({\bar{\psi}}
{\vec{\tau}}{\psi})({\bar{\psi}}{\psi}) \nonumber \\
&   & -\frac{1}{4}{{\gamma}_{S}}({\bar{\psi}}{\psi})^{4}
 -\frac{1}{4}{{\gamma}_{V}}[({\bar{\psi}}{{\gamma}_{\mu}}{\psi})({\bar{\psi
}}{{\gamma}^{\mu}}
{\psi})]^{2} \ , \ {\rm and} \\
 & &  \nonumber \\
{\cal L}_{der} & = & -\frac{1}{2}{{\delta}_{S}}({\partial_{\nu}}{\bar{\psi
}}\psi)
({\partial^{\nu}}{\bar{\psi}}\psi) \nonumber \\
&   & -\frac{1}{2}{{\delta}_{V}}({\partial_{\nu}}{\bar{\psi}}{\gamma_{\mu}}
\psi)
({\partial^{\nu}}{\bar{\psi}}{\gamma^{\mu}}\psi) \ .
\end{eqnarray}
 
\noindent And the corresponding new potentials appearing in the Dirac-Hartree
equations are given by
 
\begin{eqnarray}
{V_{S}} & = & {\alpha_{S}}{\rho_{S}} + {\beta_{S}}{\rho_{S}^{2}} + {\gamma_
{S}}{\rho_
{S}^{3}} + {\delta_{S}}{\Delta}{\rho_{S}}
+ \frac{1}{3}{\beta_{TS}}{\rho_{TS}^{2}} \ , \\
{V_{V}} & = & {\alpha_{V}}{\rho_{V}} + {\gamma_{V}}{\rho_{V}^{3}} + {\delta
_{V}}
{\Delta}{\rho_{V}} \ , \\
{V_{TS}} & = & {\alpha_{TS}}{\rho_{TS}}
+ \frac{2}{3}{\beta_{TS}}{\rho_{S}}{\rho_{TS}} \ , \ {\rm and} \\
{V_{TV}} & = & {\alpha_{TV}}{\rho_{TV}} \ .
\end{eqnarray}
 
\noindent There are now 10 coupling constants in the Lagrangian. The new
term appearing in Eq. (13) results in two additions to the potentials
appearing in the Dirac-Hartree equations, one in the isoscalar-scalar
potential, Eq. (15), that does not change sign with isospin but does contain
the isovector-scalar density $\rho_{TS}$, and one in the
isovector-scalar potential, Eq. (17), that does change sign with isospin and
also contains the isovector-scalar density. \\
 
\noindent The $\chi^{2}$ minimization process with the new Lagrangian and
10 coupling constants resulted in calculation (d) of Table 4. Eight of the
ten coupling constants are natural and the sum of the remaining two, $c_{
\gamma_{S}}$ and
$c_{\gamma_{V}}$ of the eight-fermion interaction, is also natural.
The predictive power is almost as good as that of the original NHM
Lagrangian, and is better than that of calculations (b) and (c).
However, the $\chi^{2}$/pt for this calculation is somewhat larger
than those of calculations (b) and (c) whereas one would expect
it to instead be somewhat smaller.
This behavior is at least partly due to the numerical difficulties
associated with having a {\it self-consistent} Dirac-Hartree solver as the
function call in a nonlinear least-squares minimization algorithm for
the Dirac coupling constants. \\
 
\noindent At this point the constraint $c_{\gamma_{S}}\:\equiv\:c_{\gamma_
{V}}$
was invoked and grid-search calculations were performed where the starting
values for these eight-fermion interactions were taken as one-half their sum
from calculation (b), for the original Lagrangian, and from calculation (d),
for the new Lagrangian. The results from the two grid-search calculations
are labeled (e) and (f) in \\
\newpage
\noindent Table 4. With this constraint all of the coupling
constants are natural for both the original, Eqs. (2)--(4), and the new,
Eqs. (12)--(14), Lagrangians. However, the $\chi^{2}$/pt values are factors
of $\sim$14 and $\sim$15 higher, and the predictive powers are factors of
$\sim$3 to $\sim$5 worse.
The constraint was then removed, for both Lagrangians, and full-search
calculations were performed. These resulted in small changes in the respective
sets of coupling constants; no new minima were found. \\
 
\noindent {\bf Conclusions:} Our calculations to date, summarized in Table 4,
constitute good evidence that QCD and chiral symmetry apply to finite nuclei.
The evidence, however, is at this time only partly compelling. The goal is
to construct a Lagrangian whose coupling constants are not only all
natural, but whose predictive power is superior to the original NHM
Lagrangian. This goal has not yet been achieved.
To achieve it, additional isospin dependence may be required, tensor terms
may be required, and the pions may have to be included. The work
will continue. \\
 
\noindent {\bf Acknowledgments:} I wish to thank my collaborators
B. A. Nikolaus, T. Hoch, J. L. Friar, and B. W. Lynn. This work
was performed under the auspices of the U.S.
Department of Energy. \\
 
\noindent {\bf References:}
\begin{enumerate}
\item B. A. Nikolaus, T. Hoch, and D. G. Madland, Phys. Rev. C {\bf 46}, 1757
(1992).
\item B. D. Serot and J. D. Walecka, Phys. Lett. {\bf 87B}, 172 (1979).
\item C. J. Horowitz and B. D. Serot, Nucl. Phys. {\bf A368}, 503 (1981).
\item J. L. Friar, D. G. Madland, and B. W. Lynn, Phys. Rev. C {\bf 53},
3085 (1996).
\item A. Manaohar and H. Georgi, Nucl. Phys. {\bf B234}, 189 (1984).
\item S. Weinberg, Phys. Lett. B {\bf 251}, 288 (1990).
\item S. Weinberg, Physica {\bf 96A}, 327 (1979).
\item B. W. Lynn, Nucl. Phys. {\bf B402}, 281 (1993).
\end{enumerate}
\end{document}